\documentclass[aps,prl,twocolumn,showpacs,superscriptaddress,groupedaddress]{revtex4}  
\usepackage{subfigure}
\usepackage{graphicx}  
\usepackage{dcolumn}   
\usepackage{bm}        
\usepackage{amssymb}   
\usepackage[latin1]{inputenc}
\usepackage{amsmath}
\usepackage{epstopdf}

\usepackage[hyperfootnotes]{hyperref}

\newcommand{\specificthanks}[1]{\@fnsymbol{#1}}

\newcounter{SupplCounter}
\newenvironment{Supplequation}{%
\addtocounter{equation}{-1}
\refstepcounter{SupplCounter}

\begin{equation}}
{\end{equation}}

\begin{document}

\title{Buffer-gas cooling of a single ion in a multipole radio frequency trap beyond the critical mass ratio}

\author{Bastian H{\"o}ltkemeier} \thanks{Both authors contributed equally to this work.} 
\author{Pascal Weckesser} \thanks{Both authors contributed equally to this work.}
\author{Henry L\'{o}pez-Carrera} 
\affiliation{Physikalisches Institut, Ruprecht-Karls-Universit{\"a}t Heidelberg, INF 226, 69120 Heidelberg, Germany}
\author{Matthias Weidem{\"u}ller}  \thanks{Email: weidemueller@uni-heidelberg.de}
\affiliation{Physikalisches Institut, Ruprecht-Karls-Universit{\"a}t Heidelberg, INF 226, 69120 Heidelberg, Germany}
\affiliation{Hefei National Laboratory for Physical Sciences at the Microscale and Department of Modern Physics, and CAS Center for Excellence and Synergetic Innovation Center in Quantum Information and Quantum Physics, University of Science and Technology of China, Hefei, Anhui 230026, China.}

\date{\today}

\begin{abstract}

We theoretically investigate the dynamics of a trapped ion immersed in a spatially localized buffer gas. For a homogeneous buffer gas, the ion's energy distribution reaches a stable equilibrium only if the mass of the  buffer gas atoms is below a critical value. This limitation can be overcome by using multipole traps in combination with a spatially confined buffer gas. Using a generalized model for elastic collisions of the ion with the buffer gas atoms, the ion's energy distribution is numerically determined for arbitrary buffer gas distributions and trap parameters. Three regimes characterized by the respective analytic form of the ion's equilibrium energy distribution are found. Final ion temperatures down to the millikelvin regime can be achieved by adiabatically decreasing the spatial extension of the buffer gas and the effective ion trap depth (forced sympathetic cooling).

\end{abstract}
\pacs{}
\maketitle

The ion motion inside a radio frequency (RF) trap is characterized by the interplay between a fast oscillation driven by the RF-field (micromotion) and a much slower oscillation in the confining ponderomotive potential (macromotion) \cite{Dehmelt1967}, thus representing a prototypical example of a dynamically driven nonlinear system \cite{Ghosh1996}.
Through elastic collisions with a cold buffer gas, either consisting of a cryogenic noble gas \cite{Pearson1995, schlemmer1999, glosik2006} or laser cooled atoms  \cite{Hudson2009, Grier2009, Zipkes2010a, Ravi2012, Schmid2012, Rellergert2013, Harter2013, Willitsch2014, Dutta2015}, the ion's motion can be efficiently reduced, thus opening a wide range of applications ranging from precision spectroscopy \cite{Asvany2005} and spectrometry \cite{Douglas2005, Blaum2006} over quantum computation \cite{Leibfried2004} to cold chemical reactions dynamics \cite{Mikosch2010} and astro-chemistry \cite{Gerlich1989, Gerlich2006}. However, elastic collisions influence the permanent exchange of energy between micromotion and macromotion resulting in a net energy transfer from the micromotion to the macromotion \cite{Asvany2009, Devoe2009, Zipkes2011, Cetina2012, Chen2014}. As a consequence of this coupling, the ion's final mean energy generally exceeds the buffer gas temperature \cite{Asvany2008b, Zipkes2010a, Schmid2010, Otto2012} and the ion's energy distribution is predicted to deviate from a thermal distribution \cite{Devoe2009,Zipkes2011,Chen2014}. Due to the stability constraints of a RF ion trap, efficient cooling through collisions with buffer gas atoms can only be achieved for sufficiently low atom-to-ion mass ratios $\xi = m_\mathrm{a}/m_\mathrm{i}$.
For larger mass ratios, the ion experiences an effective energy gain through elastic collisions, even if the buffer gas is at zero temperature, finally resulting in loss from the trap.
These two regimes are separated by the critical mass ratio $\xi_{\mathrm{crit}}$, as first introduced by Major and Dehmelt \cite{Dehmelt1968}.
In recent years, several groups have determined the critical mass ratio numerically \cite{Devoe2009, Zipkes2011} as well as analytically \cite{Chen2014} and found values slightly larger than Major and Dehmelt's first prediction $\xi_{\mathrm{crit}}=1$.

\begin{figure}[b]
\includegraphics[width=\columnwidth]{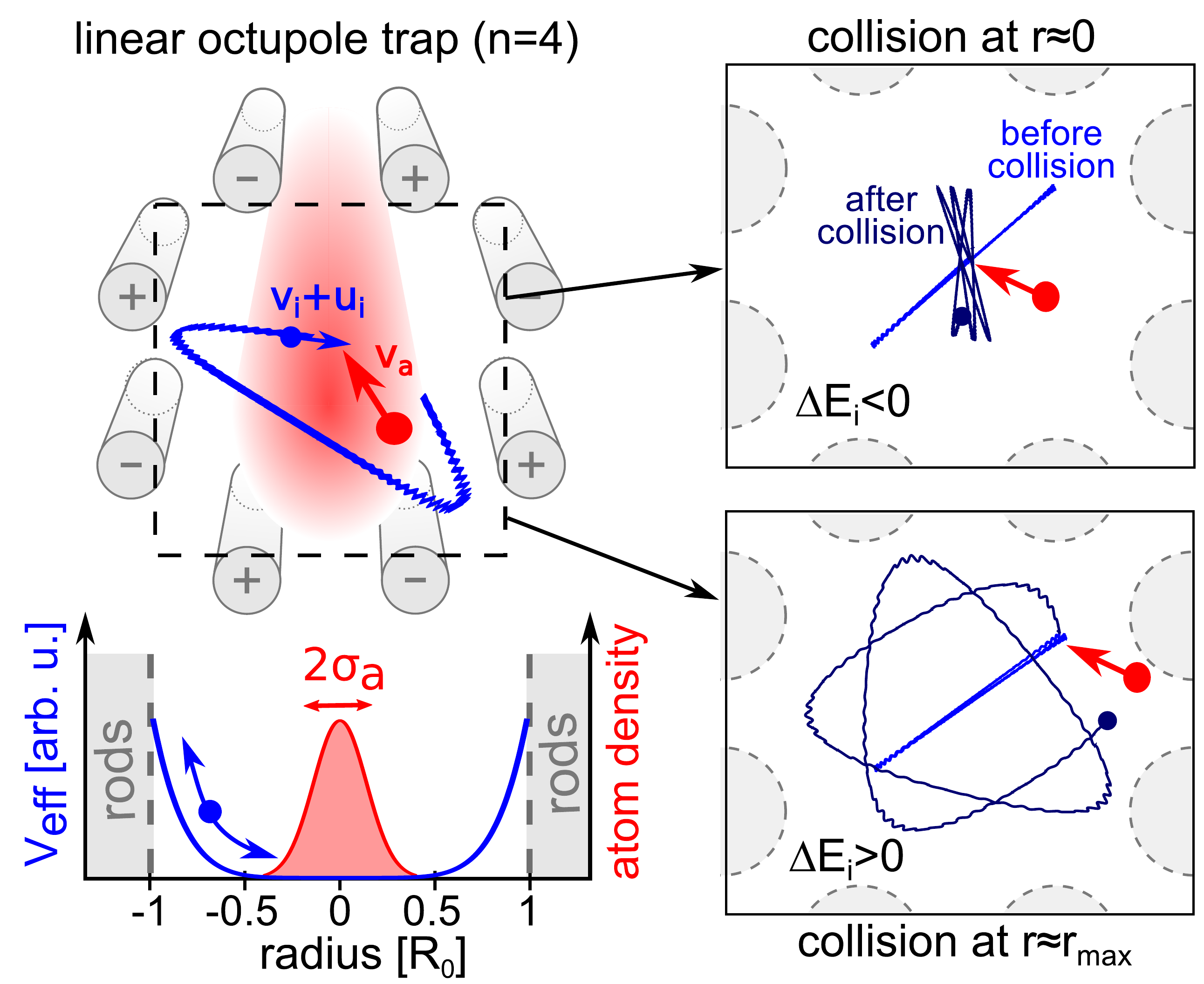}
\caption{ 
Schematic of an ion trajectory (blue) in a linear octupole trap (pole order $n=4$) colliding with an buffer gas atom (red). The velocities $\vec{v}_{\mathrm{i}}$ and $\vec{u}_{\mathrm{i}}$ are associated with the ion's micro- and macromotion, respectively, while the atom's velocity is indicated by $\vec{v}_{\mathrm{a}}$.  In the lower graph, the effective ponderomotive potential $V_\mathrm{eff}$) of the ion trap is shown together with the spatial distribution of the buffer gas atoms (spatial extension $2 \sigma_a$). In the right panels, typical ion trajectories are shown for a collision near the the trap center, yielding a decrease of the ion's energy (upper panel), and near the classical turning point, resulting in an increase of the ion's energy (lower panel).
}
\label{fig:Schematic}
\end{figure}

In order to extend the regime of efficient cooling to lower final temperatures and larger mass ratios, two different approaches have been explored.
Spatial confinement of the buffer gas, e.g. atoms stored in optical traps \cite{Grier2009, Zipkes2010a, Ravi2012, Schmid2012, Rellergert2013, Harter2013, Willitsch2014, Dutta2015}, restricts collisions to the trap center where the micromotion is smallest, thus reducing the collision induced energy transfer to the macromotion \cite{Gerlich1992, Zipkes2011, Ravi2012}.
Alternatively, in RF traps with higher pole orders the micromotion is reduced over a larger volume thus allowing for efficient cooling with buffer gas \cite{Gerlich1992, Wester2009}.
Despite a growing number of experiments using these approaches a general theoretical framework describing the influence of both a spatially confined buffer gas and high order RF traps is still lacking.

In this paper we present a comprehensive model for the dynamics of a single ion interacting with a spatially confined buffer gas inside an RF trap of arbitrary pole order.
The collisional kinematics can be favorably described in a reference frame assigning the micromotion to the buffer gas rather than the ion.
Depending on the mass ratio $\xi$, the ion's final energy is either determined by the buffer gas temperature ($\xi \ll \xi_\mathrm{crit}$) or by the effective energy of the ion's micromotion ($\xi \gg \xi_\mathrm{crit}$).
For a homogeneous buffer gas, the micromotion restricts cooling to $\xi<\xi_\mathrm{crit}$ in agreement with previous work.
However, for a spatially confined cooling agent the emergence of an additional stable regime is found, thus enabling efficient cooling of the ion motion beyond $\xi_\mathrm{crit}$.
In this regime the ion's energy distribution is determined by the energy of the trap's ponderomotive potential $V_{\mathrm{eff}}$ averaged over the buffer gas distribution. We provide semi-analytic expressions for the ion's energy distribution for arbitrary mass ratios and trap multipole orders. As the averaged ponderomotive potential can be controlled, e.g. through adjustment of the trap parameters or the atom cloud size, the ion's final temperature can actively be changed offering perspectives for enhanced cooling (\emph{forced sympathetic cooling}).

Consider an ion stored in a cylindrically symmetric RF ion trap of pole order $n$, undergoing elastic collisions with a spatially confined neutral buffer gas as schematically depicted in Fig.~\ref{fig:Schematic}.  The ion's motion can be separated into it's micromotion  and macromotion, characterized by the velocities $\vec{v}_{\mathrm{i}}$ and $\vec{u}_{\mathrm{i}}$, respectively. The micromotion is an implicit function of the ion's radial position $r$ and the RF phase $\Phi_{\mathrm{RF}}$:
\begin{equation}
\lvert \vec{v}_{\mathrm{i}} (r)\rvert = k \frac{n}{m_{\mathrm{i}}} r^{n-1} \cos{\Phi_{\mathrm{RF}}} ,
\label{eq:MiMo}
\end{equation}
with $n$ being the trap's multipole order, $m_{\mathrm{i}}$ denoting the ion's mass and $k=(N \mathrm{e^-}) V_0 / (\omega R_0^n)$ being a constant depending on the system parameters: $N \mathrm{e^-}$ (ion charge), $V_0$ (RF-voltage), $\omega$ (RF-frequency) and $R_0$ (inner trap radius). 
In this dynamically driven system the ion's energy oscillates with $\Phi_{\mathrm{RF}}$.
Generally, this oscillation is much faster than the time scale associated with the macromotion. For a Paul trap, this corresponds to a small $q$-parameter in the Matthieu equations \cite{Leibfried2004}. By averaging over one RF cycle, the ion's energy associated with the macromotion is obtained as $E_{\mathrm{i}} = m_{\mathrm{i}} \vec{u}_{\mathrm{i}}^2/2 + V_{\mathrm{eff}}(r)$ with the ponderomotive potential $V_{\mathrm{eff}} = m_{\mathrm{i}} \langle \vec{v}_{\mathrm{i}}(r)^2\rangle_{\mathrm{RF}}/2=n^2k^2r^{2n-2}/4m_{\mathrm{i}}$.

A third time scale is defined by the duration of a collision which is assumed to be short compared to the other two time scales.
This holds, as long as the collision energy exceeds the energy scale set by the interplay between micromotion and short-range atom-ion interaction, as introduced in \cite{Cetina2012}, which is typically on the order of tens or hundreds of microkelvin.
In this case, the micromotion remains unchanged during the collision ($\vec{v}_{\mathrm{i}}^{'} \approx \vec{v}_{\mathrm{i}}$), and the macromotion after a single elastic collision is given by (see Supplemental Material):
\begin{equation}
\vec{u}_{\mathrm{i}}^{'}=\frac{\xi}{1+\xi}\mathcal{R}(\theta_c,\phi_c) \cdot \left[\vec{u}_{\mathrm{i}}-(\vec{v}_a-\vec{v}_{\mathrm{i}})\right] + \frac{\vec{u}_{\mathrm{i}}+\xi(\vec{v}_a-\vec{v}_{\mathrm{i}})}{1+\xi},
\label{eq:New_MaMo}
\end{equation} 
with $\vec{v}_{\mathrm{a}}$ being the atom's velocity and $\mathcal{R}(\theta_c,\phi_c)$ being the rotation matrix defined by the polar and azimuthal scattering angles $\theta_c$ and $\phi_c$.

Note, that Eq.~\eqref{eq:New_MaMo} is formally equivalent to an elastic collision in free space of an ion with momentum $m_{\mathrm{i}} \vec{u}_{\mathrm{i}}$ colliding with an atom of momentum $m_{\mathrm{a}} (\vec{v}_{\mathrm{a}} -\vec{v}_{\mathrm{i}})$.
Therefore, it is useful to choose a reference frame in which the micromotion is assigned to the atom instead of the ion, leading to an effective atom velocity $\vec{v}_{\mathrm{eff}} = \vec{v}_{\mathrm{a}}-\vec{v}_{\mathrm{i}}$. By averaging over all collision angles and one period of the micromotion the atom's average kinetic energy is obtained as
\begin{equation}
\left< E_{\mathrm{a}}(r) \right> = \frac{1}{2}m_\mathrm{a} \left< v_{\mathrm{eff}}^2 \right> = \xi V_\mathrm{eff}(r) + \frac{3}{2} k_\mathrm{B} T_\mathrm{a}
\label{eq:effectiveEnergy}
\end{equation}
with $k_\mathrm{B}$ being the Boltzmann constant and $T_\mathrm{a}$ the buffer gas temperature.

By comparing the radial dependence of the atoms average kinetic energy to the one of the ion, as given by $\frac{1}{2}m_\mathrm{i} u_{\mathrm{i}}^2(r)$, two distinct regions, separated by a radius $r_\mathrm{c}$, can be identified: For $r<r_c$ the ion's kinetic energy exceeds the average energy of the atoms resulting in a net energy transfer to the atoms, whereas for $r>r_\mathrm{c}$ the ion's energy is generally increased through a collision with the buffer gas atom.
The radius $r_\mathrm{c}$ is given by
\begin{equation}
r_{\mathrm{c}} =r_{\mathrm{max}} \left( \frac{1- k_\mathrm{B} T_\mathrm{a}/E_\mathrm{i}}{1+\xi} \right)^{\frac{1}{2n-2}} \ ,
\label{eq:R_cool}
\end{equation}
with $r_\mathrm{max}$ being the ion's maximum turning point in the ponderomotive potential, as defined by the condition $V_{\mathrm{eff}}(r_\mathrm{max})=E_\mathrm{i}$.
For $E_\mathrm{i}< k_\mathrm{B} T_\mathrm{a}$,  the net energy transfer is always positive and, thus, the radius $r_\mathrm{c}$ no longer defined.

In order to numerically determine the ion's equilibrium energy distribution, the energy $E_\mathrm{i}$ is tracked over the course of many collisions.
Generally, this would require to solve the equations of motions for the ion's full trajectory and to evaluate the scattering probability 
at every infinitesimal time step.
In order to circumvent this computationally demanding approach, two simplifications can be made. Firstly, for the energy regime discussed here, the classical Langevin model
\cite{Langevin1905,Gioumousis1958} can be used, which results in a velocity-independent scattering probability \cite{Zipkes2011, Chen2014}.
Secondly, assuming that the time between consecutive collisions is long compared to the period of the macromotion, the scattering probability can be averaged over the macromotion resulting in an radially dependent scattering probability (see Supplemental Material).


\begin{figure}[b]
\includegraphics[width=\columnwidth, trim= 0cm 0.3cm 0cm 0.5cm]{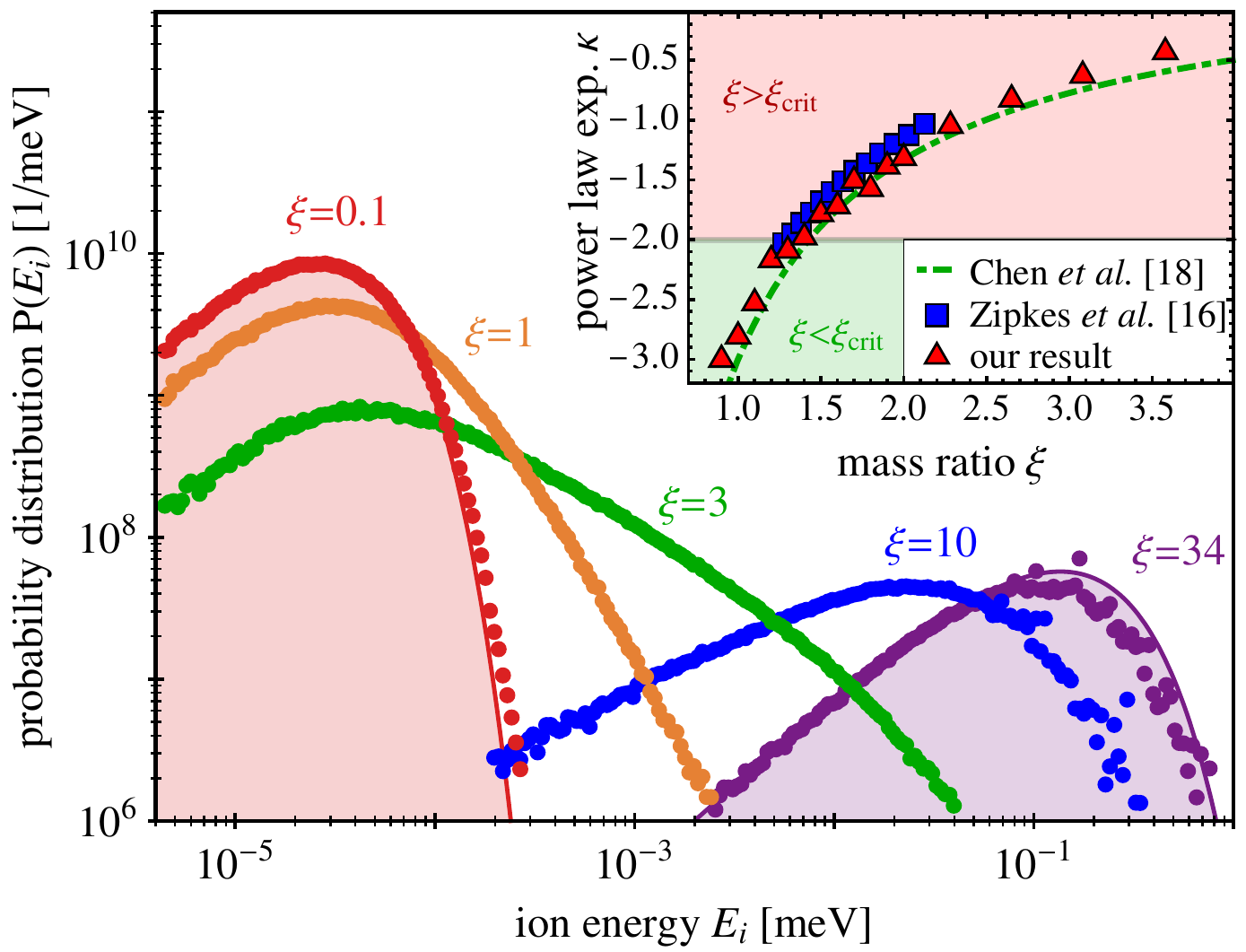}
\caption{\label{fig:EnergyDistributions}
Normalized equilibrium energy distributions for different mass ratios $\xi$ in a Paul trap (pole order $n=2$). The buffer gas cloud distribution is given by a Gaussian of size $\sigma_a=R_0/100$ and temperature of $T_a=200 \mu$K. Also shown is the energy distribution in the Boltzmann regime (red curve) and the energy distribution for $\xi = 34$ (purple curve), according to the expressions in Table~\ref{table:regimes}.
The inset compares the exponents $\kappa$ of the power-law in the energy distribution (as defined in Table~\ref{table:regimes}), for different models. The condition $\kappa = -2$ separates the regimes of stable from unstable ion motion.
Results found by Zipkes \emph{et al.} have been corrected as $\kappa_{\mathrm{corr}}=\kappa-1$ (see Supplemental Material).}
\end{figure}

Fig.~\ref{fig:EnergyDistributions} shows the resulting final distributions $P(E_{\mathrm{i}})$ for different mass ratios in a Paul trap. Depending on the mass ratio $\xi$, we find three different regimes characterized by the shape of the energy distributions.
For $\xi \ll 1$ the ion thermalizes to the buffer gas temperature as the atoms' effective energy is dominated by the atoms thermal energy  (see Eq.~\eqref{eq:effectiveEnergy}).  Therefore, the ion's thermal distribution follows a Maxwell-Boltzmann distribution for a gas in a harmonic radial potential (\emph{Boltzmann regime}, see Table~\ref{table:regimes}).

\begin{table}[b]
\begin{center}
\begin{tabular}{lll}
\hline \hline
\emph{Boltzmann regime} ($\xi\ll\xi_{\mathrm{crit}}$) & $P(E_{\mathrm{i}})\propto E_{\mathrm{i}}^{3/2} \, \exp{(-\frac{E_{\mathrm{i}}}{k_B T_a})}$ \\
\emph{Power-Law regime} ($\xi\sim\xi_{\mathrm{crit}}$) & $P(E_{\mathrm{i}})\propto \left\{
  \begin{array}{l l}
    E_{\mathrm{i}}^{3/2}, & \  E_{\mathrm{i}}\ll k_B T_a \\
    E_{\mathrm{i}}^{\kappa}, & \ E_{\mathrm{i}}\gg k_B T_a
  \end{array} \right. $ \\
	\vspace{0.1cm}
\emph{Localization regime} ($\xi\gg\xi_{\mathrm{crit}}$) & $P(E_{\mathrm{i}})\propto E_{\mathrm{i}}^{\kappa} \, \exp{(-\frac{E_{\mathrm{i}}}{E_\mathrm{a}^{\star}})}$  \\
\hline \hline
\end{tabular}
\caption{Analytical expressions for the ion's energy distribution in the three different regimes for a Paul trap, as obtained from fitting the numerical results shown in Fig.~\ref{fig:EnergyDistributions}.}
\label{table:regimes}
\end{center}
\end{table}

A second regime emerges for larger mass ratios $\xi \approx 1$.
The energy distribution exhibits a power law tail $E^{\kappa}$ towards higher energies, while the low energy behavior still follows a Boltzmann distribution (\emph{Power-Law regime}, see Table~\ref{table:regimes}).
Such power-law behavior has been identified and explained in previous investigations \cite{Asvany2009,Devoe2009,Zipkes2011,Chen2014}.
From the frame transformation , resulting in Eq.~\eqref{eq:effectiveEnergy}, the deviation from a Boltzmann distributions can be understood as a consequence of the contribution of the micromotion to the atoms' effective velocity. Through collisions with the buffer gas, an ion can gain multiples of it's current energy $V_{\mathrm{eff}}(r_{\mathrm{max}})$.

The inset of Fig.~\ref{fig:EnergyDistributions} depicts the power-law exponent $\kappa$ obtained from our simulation and compares it with previous models, showing excellent agreement. For $\kappa \geq -2$ the ion's mean energy diverges and the ion is no longer confined by the trap. This condition is commonly used to define the critical mass ratio $\xi_{\mathrm{crit}}$ \cite{Devoe2009,Chen2014}. From the simulations we obtain $\xi_{\mathrm{crit}}\approx 1.4$.

If the buffer gas is evenly distributed over the entire ion trap, sympathetic cooling is only feasible for  $\xi<\xi_{\mathrm{crit}}$. 
For an atom cloud with finite size $\sigma_\mathrm{a}$, however, we find an additional stable regime even for $\xi \gg 1$, characterized by an energy distribution following a power-law at lower energies bound by an exponential decrease towards higher energies  (\emph{Localization regime}, see Table~\ref{table:regimes}).
For increasing mass ratios, the exponent in the Power-Law regime increases, until the distribution becomes essentially flat ($\kappa \simeq 0$) for $\xi \approx 5$. For even larger mass ratios, the exponent increases further, finally converging towards $\kappa \approx 3/2$ for $\xi \gg 1$.

The exponential decrease of the energy distribution at the higher energies is caused by the localization of the buffer gas. As collisions are restricted to the volume of the buffer gas, the atoms' effective energy  $\left< E_{\mathrm{a}} (r) \right>$ (see Eq.~\eqref{eq:effectiveEnergy}) is bound by the finite size of the cloud. The effective energy content $\bar{E}_\mathrm{a}$ accessible for collisions with an ion is given by integrating $\left< E_{\mathrm{a}} (r) \right>$ over the spatial distribution of the buffer gas. For a Gaussian shaped cloud with standard deviation $\sigma_\mathrm{a}$ this yields
\begin{equation}
\bar{E}_\mathrm{a} =  \frac{3}{2} k_\mathrm{B} T_\mathrm{a} + 2^{n-1} (n-1)! \, \xi \, V_\mathrm{eff}(\sigma_\mathrm{a}) \ .
\label{eq:Emax}
\end{equation}
In the Localization regime, the scale of the energy distribution is thus given by $\bar{E}_\mathrm{a}$, in contrast to $k_\mathrm{B} T_a$ for the Boltzmann regime. 
This is illustrated in Fig.~\ref{fig:EnergyDistributions}.
For $\xi = 0.1$ the numerical data is well reproduced by a Boltzmann distribution with energy scale $k_\mathrm{B} T_a$, whereas the distribution shown for $\xi = 34$ is given by a distributions function of Boltzmann type, see  Table~\ref{table:regimes} (\emph{Localization regime}). The mean energy is given by $\bar{E}_\mathrm{a}$, therefore the characteristic energy scale $E_\mathrm{a}^{\star}$ becomes dependent on the power-law exponent $\kappa$.


So far, we have discussed the case of a Paul trap. For higher trap orders, we find the same regimes, yet a critical mass ratio $\xi_{\mathrm{crit}}(n)$ depending on the multipole-order $n$. The energy distributions are slightly modified to the ones described in Table~\ref{table:regimes}: In the Boltzmann regime the distributions differ as the average energy stored in the ponderomotive motion varies with $n$. The ratio between the ion's average potential and kinetic energy can be expressed as $\langle V_{\mathrm{eff}} \rangle / \langle E_\mathrm{kin} \rangle = 2/(3n-3)$, as follows from the virial theorem.
Thus, the ion has a larger mean energy $\langle E_{\mathrm{i}}\rangle=(\frac{3}{2}+\frac{1}{n-1})k_\mathrm{B}T_\mathrm{a}$ than the buffer gas atoms thermal energy $\frac{3}{2}k_\mathrm{B}T_\mathrm{a}$.
Therefore, the corresponding Boltzmann distribution is given by $P(E_{\mathrm{i}})\propto E_{\mathrm{i}}^{1/2 +1/(n-1)} \, \exp{(-E_{\mathrm{i}}/k_B T_a)}$, which reproduces our numerical simulations well.

With increasing mass ratio we observe the emergence of the Power-Law regime as in case of the Paul trap, with the difference that the transition now occurs at higher mass ratios. For a fixed mass ratio one finds that the exponent decreases with increasing pole order, as was already described in \cite{Asvany2009}.
The critical mass ratio, again defined by the exponent $\kappa=-2$, is numerically found to be
\begin{equation}
\xi_{\mathrm{crit}}(n) \approx 1.4 \, (n-1) \ .
\label{eq:xi_crit_numeric}
\end{equation}
The critical mass ratio can also be approximated analytically from Eq.~\eqref{eq:New_MaMo} by solving $\langle \vec{u}_{\mathrm{i}} \! ' \! \, ^2 - \vec{u}_{\mathrm{i}} \! \, ^2 \rangle_{\mathrm{RF}} = 0$ and again applying the virial theorem.
This results in a critical mass ratio of $\xi_{\mathrm{crit}}(n) = 1.5 \, (n-1)$ which reproduces the linear dependence on the pole order of the numerical findings. Consequently, Eq.~\ref{eq:xi_crit_numeric} suggests that sympathetic cooling can be efficiently applied even with a homogeneously distributed heavy buffer gases ($\xi \gg 1$) as long as the pole order $n$ is sufficiently high.

For a localized buffer gas with a mass ratio exceeding the critical value as given in Eq.~\ref{eq:xi_crit_numeric}, the ion's energy distribution follows the same analytic form as in a Paul trap (see \emph{Localization regime} in Table \ref{table:regimes}). The corresponding energy scale is given by the general expression Eq.~\ref{eq:Emax} yielding $P(E_{\mathrm{i}})\propto E_\mathrm{i}^{\kappa} \exp(-\sqrt[n-1]{E_\mathrm{i} / E_\mathrm{a}^{\star}})$. The mean energy is again given by $\bar{E}_\mathrm{a}$, therefore $E_\mathrm{a}^{\star}$ depends on $n$ as well as the power-law exponent $\kappa$.

\begin{figure}[b]
\includegraphics[width=\columnwidth, trim= 0cm 0.4cm 0cm 0.8cm]{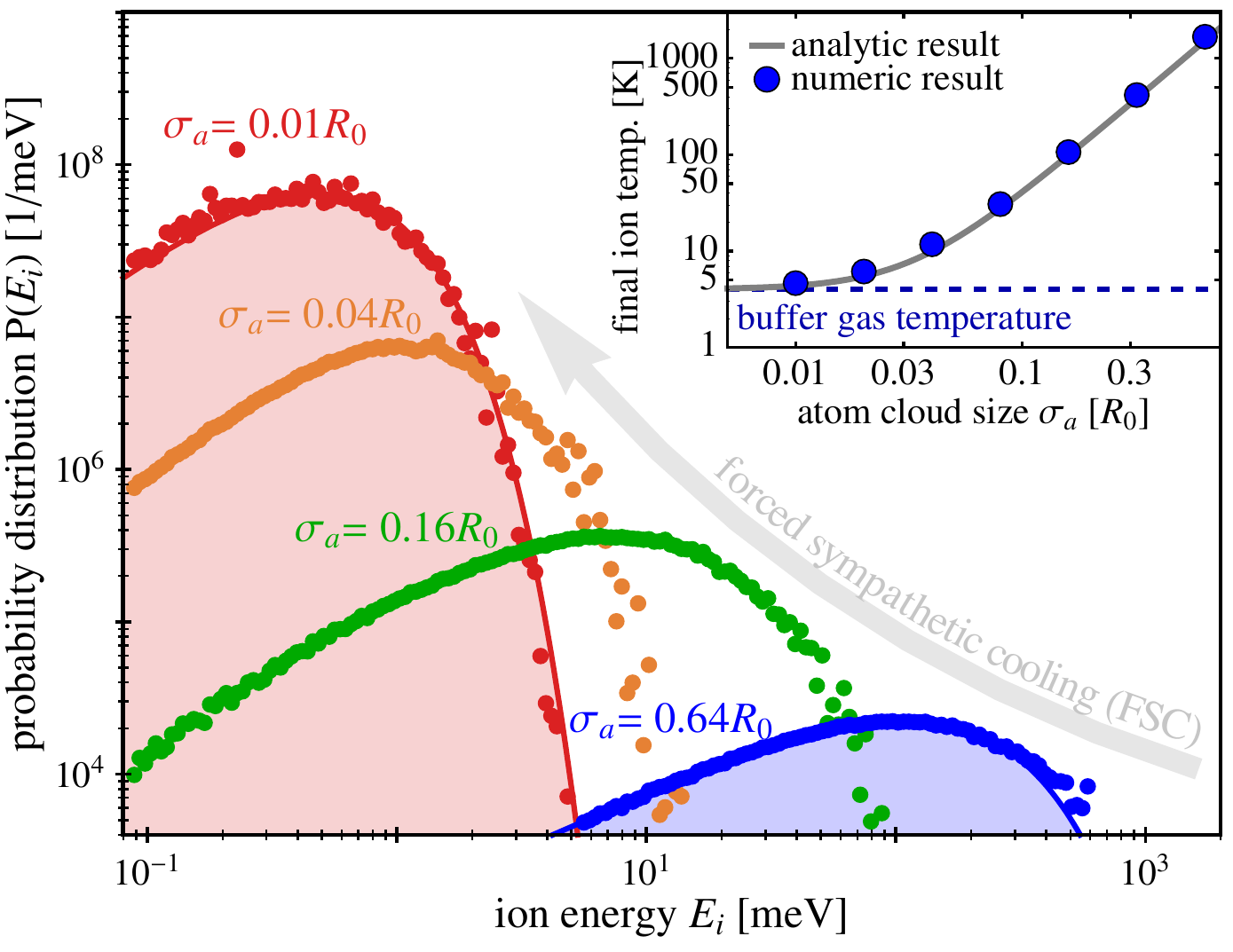}
\caption{\label{fig:FSC}
Forced Sympathetic Cooling. Shown are four energy distributions in a Paul for different buffer gas sizes.
The buffer gas has a temperature of $T_a=4$K and $\xi=10$.
The two curves correspond to the analytic expressions found in Table~\ref{table:regimes} for $\sigma_a=0.01$ and $0.64$.
The inset shows the ions final temperature as a function of the buffer gas size.
The analytic expression corresponds to Eq.~\eqref{eq:Emax}, the numeric result was obtained by fitting a modified Boltzmann distribution to the numeric data.
}
\end{figure}

The scaling of the maximum energy given by Eq.~\ref{eq:Emax} implies that the ion's mean energy can be decreased by lowering $V_\mathrm{eff}(\sigma_\mathrm{a})$ (\emph{Forced Sympathetic Cooling}, FSC).
This can be achieved by either lowering the rf voltage $V_0$, or by compressing the buffer gas cloud, or both at the same time.
Unlike forced evaporative cooling in atomic traps, FSC does not lead to a loss of particles \cite{Ketterle1996}. Adjusting these parameters sufficiently slow compared to the collisional equilibration time, the ion remains trapped.
We find in our simulations, that the ratio between the size of the buffer gas cloud and the volume probed by the ion remains constant, leaving the relative overlap and thus the average scattering rate unaffected.
This holds, as long as the thermal energy in Eq.~\ref{eq:Emax} is negligible, resulting in the ion's mean energy being proportional to $V_\mathrm{eff}(\sigma_\mathrm{a})$.
In this case the volume probed by the ion, characterized by $r_\mathrm{max}$, is directly proportional to $\sigma_\mathrm{a}$ and does not depend on the trap parameters.

Fig.~\ref{fig:FSC} demonstrates the principle of FSC by slowly decreasing the size of the buffer gas.
Following Eq. \ref{eq:Emax}, the ion's energy distribution is shifted towards lower energies.
In case of a Paul trap and a mass ratio of $\xi=10$, this leads to a decrease in the ion's temperature from 1600K for $\sigma_\mathrm{a}=0.64R_0$ down to the buffer gas temperature of 4K for $\sigma_\mathrm{a}=0.01R_0$ (see inset of Fig.~\ref{fig:FSC}).

Using the favorable frame transformation, where the micromotion is assigned to the buffer gas atoms, we have found simple analytic expressions describing the energy distribution of the ion and provided intuitive insight into the dynamics of buffer-gas cooling in ion traps of arbitrary multipole order. For a spatially confined buffer gas, we discovered a regime of stable ion motion even for large atom-to-ion mass ratios, thus extending previous investigations on the dynamics of ions colliding with a homogeneously distributed buffer gas. These findings enable the design of a new class of experiments for creating cold and ultracold samples of atomic and molecular ions. By ramping down the ion trap voltage and/or the spatial extension of the buffer gas, one can further cool the ion's motion (Forced Sympathetic Cooling). As a realistic example \cite{Deiglmayr2012} we estimate a final temperature of a few tens of millikelvin for OH$^-$ ions in an octupole trap cooled by Rb atoms in a magneto-optical trap despite the mass ratio of $\xi=5$. So far, we have ignored ion-ion interactions \cite{Chen2013} as well as the axial confinement \cite{Otto2012, Lakhmanskaya2014} or the effect of stray electric fields \cite{Zipkes2011}.
To a first approximation these additional factors might be treated as perturbations of the ponderomotive potential, resulting in a reduction of the critical mass ratio.

This work is supported in part by the Heidelberg Center for Quantum Dynamics and the BMBF under contract number 05P12VHFA6.
B.H. acknowledges support by HGSHire and P.W. by Deutschlandstipendium and Springer.
We thank S. Whitlock, J. Evers, R. Wester and P. Schmelcher for fruitful discussions.

\bibliographystyle{plain}



\section{Appendix A: Collisions in an rf trap - derivation of Eq. (2)}
\label{Appendix_A}

Consider a single elastic collision between an atom and ion inside an rf trap with $m_{\mathrm{a}}/m_{\mathrm{i}}$ and $\vec{v}_{\mathrm{a}}/\vec{c}_{\mathrm{i}}$ being their mass and initial velocity in the lab frame, respectively.
All initial lab frame velocities can be expressed by the center-of-mass (COM) $\vec{V}= (m_{\mathrm{i}}\vec{c}_{\mathrm{i}}+m_{\mathrm{a}}\vec{v}_{\mathrm{a}})/(m_{\mathrm{i}}+m_{\mathrm{a}})$ and the relative velocity $\vec{v}_{\mathrm{0}}=\vec{c}_{\mathrm{i}}-\vec{v}_{\mathrm{a}}$: ($\vec{c}_{\mathrm{i}}=\vec{V}+\frac{\mu}{m_{\mathrm{i}}}\vec{v}_{\mathrm{0}},\vec{v}_{\mathrm{a}}=\vec{V}+\frac{\mu}{m_{\mathrm{a}}}\vec{v}_{\mathrm{0}}$).
Here $\mu=m_{\mathrm{i}}m_{\mathrm{a}}/(m_{\mathrm{i}}+m_{\mathrm{a}})$ is the reduced mass.
Elastic collisions rotate the relative velocity in the COM frame, without changing its' magnitude: $\vec{v}_{\mathrm{0}} \! '=\mathcal{R}(\theta_c,\phi_c) \vec{v}_{\mathrm{0}}$, with $\theta_0$ and $\phi_0$ being the polar and the azimuthal scattering angle, respectively.
The final velocity of the ion in the lab frame is obtained by adding the COM motion $\vec{V}$ onto the rotated relative velocity:
\begin{Supplequation}
\vec{c}_{\mathrm{i}} \! '=\frac{m_{\mathrm{a}}}{m_{\mathrm{i}}+m_{\mathrm{a}}}\mathcal{R}(\theta_c,\phi_c) (\vec{c}_{\mathrm{i}}-\vec{v}_{\mathrm{a}})+ \frac{m_{\mathrm{i}}\vec{c}_{\mathrm{i}}+m_{\mathrm{a}}\vec{v}_{\mathrm{a}}}{m_{\mathrm{i}}+m_{\mathrm{a}}} 
\label{eq:elastic_collision}
\end{Supplequation} 
In the rf trap the ion's motion can be decomposed into the macro- and the micromotion: $\vec{c}_{\mathrm{i}}=\vec{u}_{\mathrm{i}}+\vec{v}_{\mathrm{i}}$.
The micromotion is a well-defined velocity only depending on the radial distance to the trap center $r$ and the rf phase $\Phi_{\mathrm{RF}}$.
As the duration of the collision is typically small compared to the period of the micromotion, one assumes that the micromotion remains unchanged during the course of the collision, resulting  in $\vec{c}_{\mathrm{i}}  \! '=\vec{u}_{\mathrm{i}} \! '+\vec{v}_{\mathrm{i}}$.
The ion's energy is given by $E_\mathrm{i}=\frac{1}{2}m_i\vec{u}^2_{\mathrm{i}}+V_{\mathrm{eff}}$.
The energy difference is given by: $\Delta E_\mathrm{i}=\frac{1}{2}m_i (\vec{u}_{\mathrm{i}} \! ' \! \, ^2 -\vec{u}_{\mathrm{i}})$.
The final macromotion velocity $\vec{u}_{\mathrm{i}} \! '$ is obtained by $\vec{c}_{\mathrm{i}} \! ' -\vec{v}_{\mathrm{i}}$ which leads to:
\begin{Supplequation}
\vec{u}_\mathrm{i} \! ' = \frac{m_{\mathrm{a}}}{m_{\mathrm{i}}+m_{\mathrm{a}}}\mathcal{R}(\theta_c,\phi_c) (\vec{u}_{\mathrm{i}}-(\vec{v}_a-\vec{v}_{\mathrm{i}}))+ \frac{m_{\mathrm{i}}\vec{u}_{\mathrm{i}}+m_{\mathrm{a}}(\vec{v}_{\mathrm{a}}-\vec{v}_{\mathrm{i}})}{m_{\mathrm{i}}+m_{\mathrm{a}}}
\label{eq:elastic_collision_in_rf}
\end{Supplequation}
One should note that by choosing $\vec{c}_\mathrm{i} = \vec{u}_\mathrm{i}$ and $\vec{v}_\mathrm{a} = \vec{v}_\mathrm{a} - \vec{v}_\mathrm{i}$, Eq.~\eqref{eq:elastic_collision} can be converted into Eq.~\eqref{eq:elastic_collision_in_rf}.
Consequently, an atom-ion collision in the driven field of an rf trap can be described as a collision in free space of an ion and atom with modified momenta $m_{\mathrm{i}}\vec{u}_{\mathrm{i}}$ and $m_{\mathrm{a}}(\vec{v}_\mathrm{a}-\vec{v}_\mathrm{i})$.

\section{Appendix B: Numerical simulation of the energy distribution}
\label{Appendix_B}

\begin{figure}[b]
\includegraphics[width=\columnwidth, trim= 0.3cm 0.2cm 2cm 0.5cm]{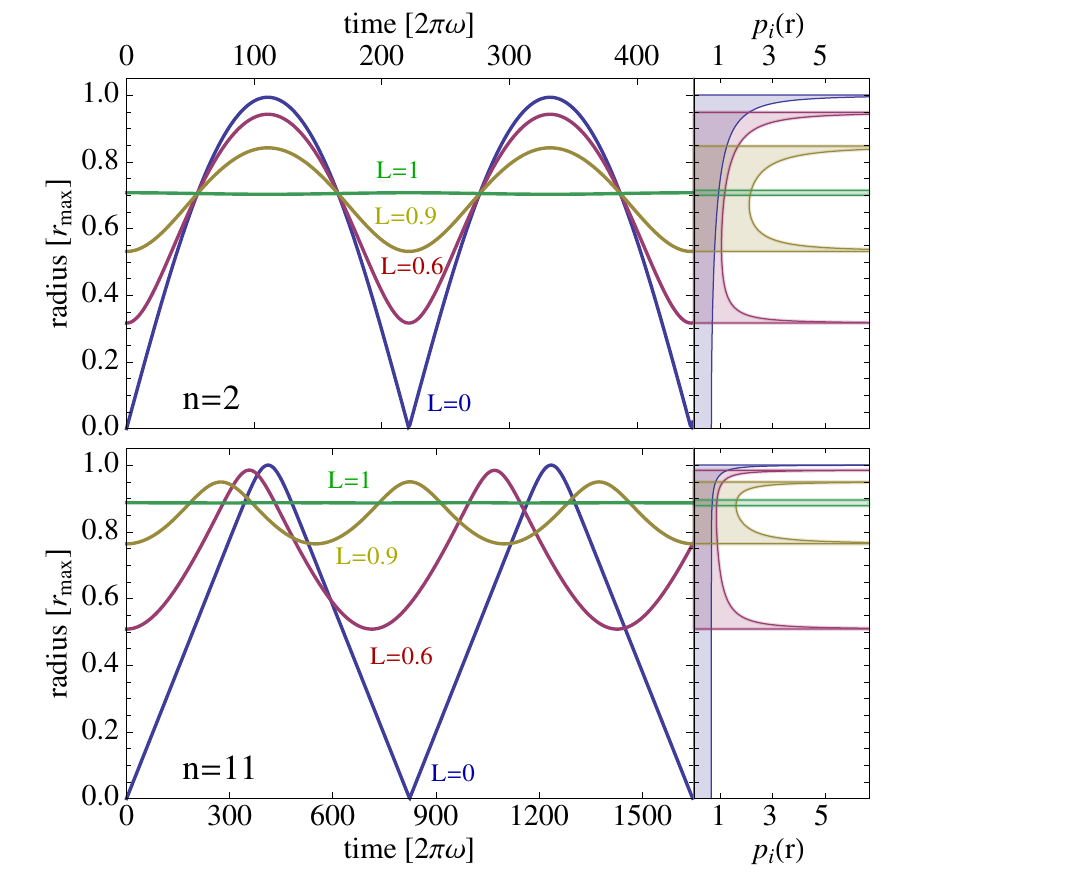}
\caption{Ion's radial probability distribution for different angular momenta. All angular momenta are expressed in terms of the the maximum angular momentum. The case $L=0$ corresponds to a trajectory through the trap center, the case $L=1$ corresponds to a circular orbit leading to $r(t)=\mathrm{const}$. The left graphs show the ion's trajectory $r(t)$, the right graphs show the probability distribution $p_i(r)$ to find the ion in an interval d$r$ around $r$. The upper graphs correspond to n=2, the lower graphs to $n=22$.
}
\label{fig:radialProb}
\end{figure}
In order to numerically calculate the ion's energy distribution, the ion is tracked over the course of typically $10^6$ collisions.
This is commonly done by calculating the ion's exact trajectory and evaluating the collision probability $P(t) \mathrm{d}t$ for every infinitesimal time step $\mathrm{d}t$.
This approach leads to very long computation times, as the time between two collisions is usually much longer than the time scales set by micro- and macromotion.
In order to circumvent this problem, the large difference of the time scales can be used to separate them.
As long as the ion undergoes many oscillations in the ponderomotive potential (macromotion) between every two consecutive collisions, the ion's exact trajectory can be substituted by a radial density distribution $p_\mathrm{i}(r)$.
The density distribution is independent of the other two cylindrical coordinates $z$ and $\phi$, as we assume perfect symmetry in these dimensions.
In this case, the probability of an collision to occur in the interval $dr$ around a radial position $r$ is given by $P(r)\,dr=\rho_\mathrm{a}(r) \sigma v_{\mathrm{rel}} p_i(r) \,dr$, with $\rho_\mathrm{a}(r)$ being the atoms' density distribution, $\sigma$ being the collision cross section and $v_{\mathrm{rel}} = \vert \vec{u}_\mathrm{i} - \vec{v}_\mathrm{eff} \vert$ being the relative atom-ion velocity.
Further assuming, that the cross section is given by the Langevin expression $\sigma \propto 1/v_{\mathrm{rel}}$, the relative velocity drops out of the scattering probability and $P(r)$ is proportional to the overlap between the atoms' density distribution and ion's spatial probability distribution.

The ion's spatial probability distribution $p_\mathrm{i}(r)$ is characterized by the ion's total energy and angular momentum.
The total energy can be separated into two components, the energy $E_\mathrm{z} = m_\mathrm{i} u_\mathrm{i,z}^2 /2$ in axial direction without confining potential and the radial energy $E_\mathrm{r} = m_\mathrm{i} u_\mathrm{i,r}^2 /2 + V_\mathrm{eff}(r)$ consisting of macromotion and ponderomotive potential, with $u_\mathrm{i,r}$ and $u_\mathrm{i,z}$ being the radial and axial components of the macromotion.
Together with the ion's angular momentum $L= m_\mathrm{i} \vert \vec{r} \times \vec{u}_\mathrm{i,r} \vert$, these quantities are constants of the ion's motion and uniquely define $p_\mathrm{i}(r)$.
The probability distribution is proportional to the inverse of the ion's radial velocity $\dot{r}$, which can be expressed as
\begin{Supplequation}
p_\mathrm{i}(r) \propto \left(\frac{2}{m_\mathrm{i}} (E_\mathrm{r}-V_\mathrm{eff}(r)) - \frac{L^2}{m_\mathrm{i}^2 r^2} \right)^{-1/2} \, .
\label{eq:bob}
\end{Supplequation}
Fig.~\ref{fig:radialProb} shows the resulting radial probability distribution $p_i(r)$ for different angular momenta.

In the simulation, the ion's radial and axial energy as well as it's angular momentum are computed and stored after every collision.
As initial conditions we typically use $E_\mathrm{r}=k_\mathrm{B} T_\mathrm{a}$, $E_\mathrm{z}=k_\mathrm{B} T_\mathrm{a} /2$ and $L=0$.  
For every collision we then pick the following set of parameters:
\begin{itemize}
\item \textbf{RF-phase} - the phase of the rf-field $\Phi_\mathrm{RF}$ is randomly chosen between 0 and 2$\pi$.
\item \textbf{Atom velocity} - all three cartesian coordinates of the atom's velocity vector are chosen with a normal distribution with standard deviation $\sqrt{k_\mathrm{B} T_\mathrm{a}/m_\mathrm{a}}$.
\item \textbf{Collision radius} - the radial position of the collision $r_\mathrm{coll}$ is chosen according to $P(r)$. For large radial ion energies P(r) can have a long tail of near zero values, in which case we use an upper boundary of $r_\mathrm{coll} < 5 \sigma_\mathrm{a}$.
\item \textbf{Scattering angles} - using the Langevin model, the scattering angle in the center-of-mass frame is distributed isotropically. This is achieved by randomly choosing an azimuthal angle $\phi_\mathrm{c}$ between zero and $2\pi$ and a polar angle $\theta_\mathrm{c}$ between zero and $\pi$ taking into account the Jakobian determinant, yielding a probability distribution $p(\theta_\mathrm{c})=\sin{\theta_\mathrm{c}}$.
\end{itemize}
Based on these parameters we can calculate the ions micro- and macromotion velocities. 
Plugging all of these parameters into Eq.~\eqref{eq:elastic_collision_in_rf}, the ion's macromotion after the collision is obtained which defines the ion's radial and axial energy as well as it's angular momentum after the collision.
As a last step we compute the average time of free motion at this energy and angular momentum shell as given by the inverse of the overlap of atom and ion distribution $\tau = (\int \rho_\mathrm{a}(r) p_i(r) \mathrm{d}r)^{-1}$.
After performing $10^6$ collisions all energy values are weighted with the corresponding $\tau$ and binned, resulting in the final energy distribution.

\section{Appendix C: Scaling the power-laws from Zipkes \textit{et al.}  \cite{Zipkes2011}}
\label{Appendix_C}

Zipkes \textit{et al.} \cite{Zipkes2011} determine the steady state solution of a single ion immersed in an ultracold atomic cloud.
In their model the ion undergoes $10^8$ consecutive collisions which are recorded and afterwards binned on a logarithmic scale, i.e. choosing bins with a logarithmic distribution of sizes.
In this way, the logarithmic energy probability distribution $dP(E_{\mathrm{i}})/d\log(E_{\mathrm{i}})$ is not properly normalized.
The logarithmic distribution function relates to the normalized linear distribution $dP(E_{\mathrm{i}})/dE_{\mathrm{i}}$ by
\begin{Supplequation}
dP(E_{\mathrm{i}})/d\log(E_{\mathrm{i}}) \propto E_{\mathrm{i}} \cdot dP(E_{\mathrm{i}})/dE_{\mathrm{i}}.
\label{eq:mapping}
\end{Supplequation}
If this operation is applied to a power-law distribution, the resulting power-law coefficient of the normalized distribution is reduced by unity. Therefore, the coefficient of the power law $\kappa$ found in Ref. \cite{Zipkes2011} has been corrected as $\kappa_\mathrm{core} = \kappa -1$ resulting in excellent agreement with the other models, including ours.

%

\end{document}